\documentclass[aps,twocolumn,showpacs]{revtex4}
\usepackage{graphicx,amsmath,hyperref} 
\begin{document}
\title{Scale-free topology of e-mail networks}
\author{Holger Ebel}
\email{ebel@theo-physik.uni-kiel.de}
\author{Lutz-Ingo Mielsch}
\author{Stefan Bornholdt}
\affiliation{Institut f\"ur Theoretische Physik, Universit\"at Kiel,
Leibnizstra\ss{}e 15, D-24098 Kiel, Germany}

\date{February 11, 2002}

\begin{abstract}
We study the topology of e-mail networks with e-mail addresses as nodes and e-mails as links using data from server log files. The resulting network exhibits a scale-free link distribution and pronounced small-world behavior, as observed in other social networks. These observations imply that the spreading of e-mail viruses is greatly facilitated in real e-mail networks compared to random architectures.
\end{abstract}

\maketitle

Complex networks as the World Wide Web or social networks often do not have an engineered architecture but instead are self-organized by the actions of a large number of individuals. 
From these local interactions non-trivial global phenomena can emerge as, for example, small-world properties \cite{watts/strogatz:1998} or a scale-free distribution of the degree \cite{barabasi/albert:1999}. These global properties have considerable implications on the behavior of the network under error or attack \cite{albert/barabasi:2000b}, as well as on the spreading of information or epidemics \cite{pastor-satorras/vespignani:2001}.
Here we report that networks composed of persons connected by exchanged e-mails show both the characteristics of small-world networks and scale-free networks.

\begin{figure}
\includegraphics[width=5.5cm,angle=-90]{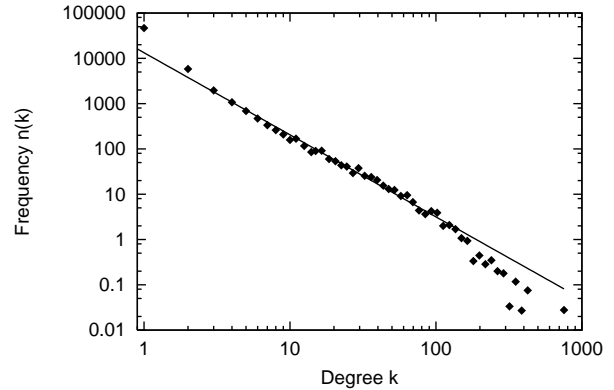} 
\caption{\label{fig_deg_all}Degree distribution of the e-mail network. The double-logarithmic plot of the number of e-mail addresses with which a node exchanged e-mails exhibits a power-law with exponent -1.81 $\pm$ 0.10 over two decades. This distribution is used to calculate estimates for the clustering coefficient and the average shortest path length for the entire network (see text).}
\end{figure}

The nodes of an e-mail network correspond to e-mail addresses which are connected by a link if an e-mail has been exchanged between them. The network studied here is constructed from log files of the e-mail server at Kiel University, recording the source and destination of every e-mail from or to a student account over a period of 112 days \footnote{Three e-mail addresses has been excluded as artifacts, since they reached a large quantity of the students only because all students used the same server (e.g.\ by service e-mails from the university computer center). Therefore their large degree was caused solely by sampling log files from only one e-mail server.}.
The resulting network consists of $N=$ 59,912 nodes (including 5,165 student accounts) with a mean degree of $<k>$ = 2.88 and contains several separated clusters with less than 150 nodes and one giant component of 56,969 nodes (mean degree $<k_{\text{large}}>$ = 2.96). The degree distribution $n(k)$, i.e.\ the distribution of the number $k$ of a node's next neighbors, obeys a power law
\begin{equation}
n(k) \propto k^{-1.81},
\end{equation}
with exponential cut-off (Fig.\ \ref{fig_deg_all}).

Most of the scaling exponents reported so far for the degree distributions of computer and social networks lie in the range of -2.0 to -3.4 \cite{albert/barabasi:2002}. One exception is the social network of co-authorships in high energy physics, for which Newman found an exceptionally small scaling exponent of -1.2 \cite{newman:2001b}. 
Similar to our work are studies of networks of phone calls made during one day. These phone-call networks show scale-free behavior of the degree distribution as well, with an exponent of -2.1 \cite{abello/resende:1999b,aiello/lu:2000}.

\begin{figure}
\includegraphics[width=5.5cm,angle=-90]{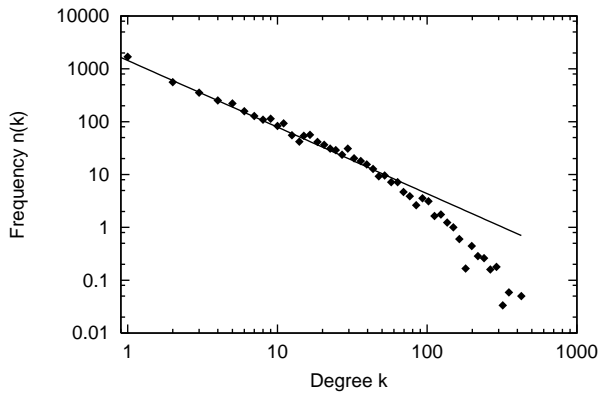}
\caption{\label{fig_deg_stu}Degree distribution of the student accounts in the e-mail network. The degree distribution of the subset of student e-mail addresses with completely known degree can be approximated by a power law as well with exponent -1.32 $\pm$ 0.18. This exponent is smaller than for the whole network since the degree of external nodes is underestimated by the measurement.}
\end{figure}

Let us briefly  discuss how our result on e-mail networks may be influenced by the measurement process. The sampling of the network has been restricted to one distinct e-mail server. Therefore, only the degrees of accounts at this server are known exactly. Here, these internal accounts correspond to e-mail addresses of local students, whereas the external nodes are given by all other e-mail addresses. We resolve the degree distribution of internal accounts only (Fig.\ \ref{fig_deg_stu}), and find that it can be approximated by a power-law $n_{\text{int}}(k) \propto k^{-1.32}$ as well (mean degree $<k_{\text{int}}>=$ 14.86). Since the degrees of external nodes typically are underestimated, this exponent is smaller than for the whole network. For the same reason, there are fewer nodes with small degree in the distribution of students' degrees. Note that the cut-off of both distributions is about the same. Therefore, external sources addressing almost all internal nodes (e.g.\ advertisement or spam) do not bias the degree statistics. 
Thus, it can be concluded that the e-mail network exhibits scale-free behavior.

Furthermore, the e-mail network shows the properties of a ``small world''  \cite{watts/strogatz:1998}, i.e.\ a high probability that two neighbors of one node are connected themselves (clustering) and a small average length $\ell$ of the shortest path between two nodes.
The clustering is measured 
by the clustering coefficient $C$ of a network which is defined in the following way: 
The clustering coefficient $C_\nu$ of a node $\nu$ 
is given by the ratio of existing links $E_\nu$ between its 
$k_\nu$ first neighbors to the potential number of such ties 
$\frac{1}{2} k_\nu (k_\nu-1)$.
By averaging $C_\nu$ over all nodes one arrives at  
the clustering coefficient 
$C$ of the network
\begin{equation}
C = \langle C_\nu \rangle_\nu = \left \langle \frac{2 E_\nu}{k_\nu (k_\nu-1)} 
\right \rangle_\nu.
\label{eq_c}
\end{equation}

To verify the small-world properties, experimental results are compared to the respective values derived for random networks, as well as for networks with links assigned randomly but obeying the same degree distribution.
We will call a network a ``random network'' if the probability $p$ that an arbitrary pair of nodes is connected is constant $p=\langle k \rangle 
/ (N-1)$, leading to a Poissonian degree distribution. In this case, the clustering coefficient is just this probability $C_{\text{rand}}=p$.
Additionally, we deduced an estimate $C'$ for an upper bound
of the 
clustering coefficient of a network with identical degree 
distribution, but randomly assigned links. Hence, $C'$ gives 
an upper bound of the clustering that is expected
from the degree distribution alone. Employing the generating function approach for networks with arbitrary degree distributions \cite{newman/watts:2001} and assuming 
that fluctuations of the mean degree of a node's neighborhood 
can be ignored, results in:
\begin{equation}
C' = \frac{1}{\langle k \rangle N} 
\left (\frac{\langle k^2 \rangle}{\langle k \rangle}-1\right 
)^2.\label{upper_bound}
\end{equation}
This equation exactly applies in the case of the Poissonian degree
distribution of a random network ($C_{\text{rand}} = C'$).
Next we compare the experimental average shortest path length 
$\ell$ with the respective value $\ell'$ of a network with
 the same degree distribution and randomly assigned links.
With the generating function approach we obtain \cite{davidsen/bornholdt:2002}:  
\begin{equation}
\ell' \approx \frac{\log \left (\frac{N}{\langle k \rangle}
\right )}{\log \left (\frac{\langle
 k^2 \rangle - \langle k \rangle}{\langle k \rangle}\right )}+1.\label{ellprime}
\end{equation}
For the Poissonian distribution of a random network, Eqn.\ (\ref{ellprime}) simplifies to:
\begin{equation}
\ell_{\text{rand}} \approx  \frac{\log N}{\log \langle k \rangle}.
\end{equation}
Note that only $N$, $\langle k \rangle$ and $\langle k^2 \rangle$ are 
used for the estimates $\ell'$ and $\ell_{\text{rand}}$ as in the derivation of 
Eqn.\ (\ref{upper_bound}).

In the following, the experimental values are compared to the above estimates. The experimental clustering coefficient $C =$ 0.156 is about one magnitude larger than one would expect solely from the degree distribution ($C'$ = 0.0187). For a random network, the respective clustering coefficient is $C_{\text{rand}}=4.82 \cdot 10^{-5}$.

The mean shortest path length in the giant component was determined to $\ell$ = 4.95 $\pm$ 0.03 with the Dijkstra algorithm \cite{rosen:2000}. It is larger than the mean shortest path length in a network with the same degree distribution \footnote{To be specific, $\ell$ has to be calculated using the degree distribution of the giant component. Since this distribution does not differ significantly from the degree distribution of the whole network, employing the latter yields the same result within numerical accuracy used here.} ($\ell'$ = 3.43) since more links are consumed for forming local clusters. It is still smaller than the path length of a random network $\ell_{\text{rand}}$ = 10.10 (where each pair of nodes is connected with a constant probability leading to the same mean degree) because of the highly connected "hubs" present in a scale-free network.

\begin{figure}[b]
\includegraphics[width=5.5cm,angle=-90]{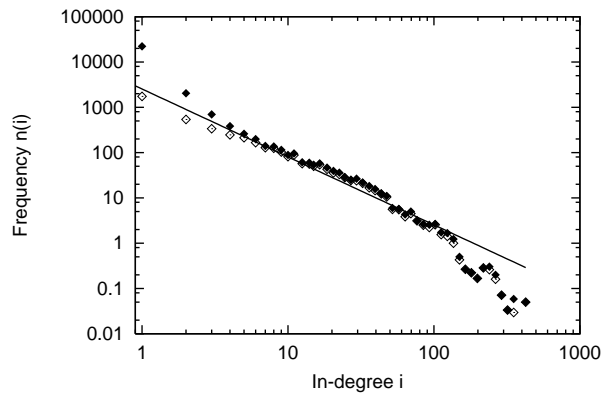}
\caption{\label{fig_deg_in} In-degree distributions for the e-mail network. The double-logarithmic plot of the in-degree distributions for all nodes (filled diamonds,$<k_{\text{in}}>=$1.62) and for student nodes only (open diamonds, $<k_{\text{in}}>_{\text{int}}=$13.06) shows a power-law distribution  with an exponent of -1.49 $\pm$ 0.18. Note that again the in-degree of external nodes is underestimated by the measurement process.}
\end{figure}

\begin{figure}
\includegraphics[width=5.5cm,angle=-90]{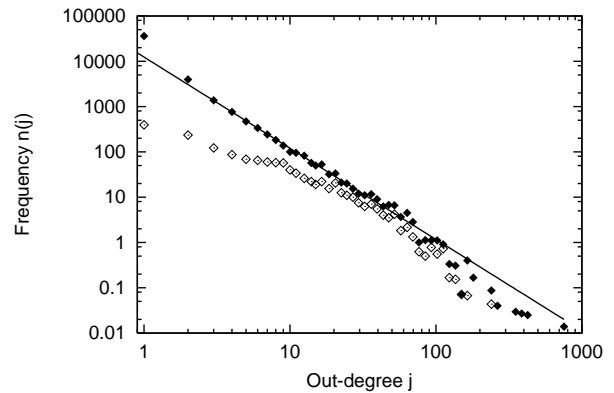}
\caption{\label{fig_deg_out} Out-degree distributions for the e-mail network. For the out-degree, only the distribution of internal and external nodes (filled diamonds, $<k_{\text{out}}>=$1.62) exhibits a pronounced power-law over two decades with exponent -2.03 $\pm$ 0.12. The distribution of the out-degree of internal nodes (open diamonds, $<k_{\text{out}}>_{\text{int}}=$12.39) is broad as well but cannot be identified with a scale-free regime which may be due to the limited size of the sample.}
\end{figure}

To further investigate the emergence of the scale-free degree distribution, we study the e-mail network as a directed graph, where an e-mail corresponds to a directed link pointing from the sender to the receiver. Although the e-mail network has to be treated as an undirected graph in the context of virus spreading (see below), it seems reasonable that the sending and receiving of e-mails are governed by different processes. Again, the analysis is done for the distributions of all nodes and of internal nodes only, where for the latter, the in- and out-degree can be determined exactly. The distribution of the in-degree $i$, i.e.\ a node's number of different nodes it has received e-mails from, are very similar for all nodes and for internal nodes, respectively (Fig.\ \ref{fig_deg_in}). They can both be approximated by a power law $n(i) \propto i^{-1.49}$ over about two orders of magnitude. Deviations of the two distributions for in-degrees $i < 6$ are due to the underestimation of the degree of external nodes. One explanation for an in-degree exponent of about -1.5 is the assumption of stochastic multiplicative growth as in the model of Huberman and Adamic \cite{huberman/adamic:1999,adamic/huberman:2000}. They proposed that the number of links a node received at a time step is a random fraction of the number of links it already has received.
The treatment of the out-degree is more difficult. For the whole network, the distribution of out-degree $j$, i.e.\ a node's number of links to other nodes, shows pronounced scale-free behavior $n(j) \propto j^{-2.03}$ (Fig. \ref{fig_deg_out}). However, the corresponding distribution for internal nodes is broad but does not show scale-free behavior over a sufficient range. This may be caused by the limited size of the sample but may also point to the systematic error caused by students possibly using different (external) accounts for sending e-mails. The out-degree scaling exponent of the whole network lies in a quite common range for communication and social networks, as, e.g., the movie actors' network or the phone call network \cite{albert/barabasi:2002}, where the principle of {\em preferential attachment} can be used for modeling \cite{barabasi/albert:1999}. It applies to the assumption that the probability $p_j$ that a link originates in the set of nodes with out-degree $j$ is proportional to the number of links already starting in this set $f[j]$:
\begin{equation}
p_j \propto j f[j].
\end{equation}
This corresponds to Simon's general model for such copy and growth processes \cite{simon:1955,bornholdt/ebel:2001}. Let us briefly apply this model to the e-mail network. From our data we estimated the ratio of the growth rate of nodes to the growth rate of links to $\alpha=$0.597 nodes per links which is sufficient to calculate the scaling exponent $\gamma$ \cite{bornholdt/ebel:2001}:
\begin{equation}
\gamma = 1 + \frac{1}{1-\alpha}.
\end{equation}
Thus, the preferential linking model leads to a steep exponent of -3.48 not in accordance with observation. On the other hand, a model based only on {\em transitive linking} \cite{davidsen/bornholdt:2002}, i.e.\ on the assumption that two nodes are more likely to be linked if they have a common neighbor, can in principle reproduce the small-world properties and a broad degree distribution but leads to a too high clustering and does not yield a power-law degree distribution for this particular network. From this we conjecture, that including both preferential and transitive linking may consistently model the e-mail network.

What are the implications of the above results for the spreading of e-mail viruses?
The occurrence of e-mail viruses has become a well-known phenomenon in today's communication experience. An e-mail virus or e-mail worm is a program attached to an e-mail which, when opened by the recipient, causes the recipient's e-mail program to remail numerous infected e-mails to e-mail addresses found in the address book or in stored e-mails. Hence, for the propagation of e-mail viruses the network is undirected. This is different for chain e-mails, where each recipient is asked to forward the chain e-mail to other addresses. E-mail viruses can cause serious damage to computer networks by destroying data at infected computers or by overloading e-mail servers and other infrastructure. In May 2000, for instance, the e-mail worm ``I love you'' infected more than 500,000 individual systems worldwide \cite{certccusa:2000}
and obstructed 21 \%
of the computer workplaces in Germany \cite{bmi:2001}.

In scale-free networks, the threshold for the propagation rate above which an infection of the network spreads and becomes persistent is very much lower than in other disordered networks, or even vanishes \cite{pastor-satorras/vespignani:2001b}. 
This means that the self-organized structure of the e-mail network facilitates the spreading of computer viruses, as well as of any other information. In addition, the e-mail network is quite robust in case of ``failures'' of random nodes if, for instance, some participant does not answer e-mails for a while or uses anti-virus software. However, it is sensitive to the loss of highly connected participants because of the power-law degree statistics \cite{albert/barabasi:2000b}. 
Hence uniformly applied immunization of nodes is less likely to eradicate infections until almost all participants are immunized, whereas targeting prevention efforts at the highly connected sites significantly suppresses epidemic outbreaks and prevalence \cite{pastor-satorras/vespignani:2001c,newman:2002}.

These observations suggest helpful and advantageous applications, but also point to the inherent dangers of e-mail networks. The security of e-mail communication can be improved by identifying highly connected hub addresses and monitoring them for viruses more strictly, e.g.\ in corporate e-mail networks to prevent the damaging and costly spreading of e-mail viruses. In a different application, making use of the high clustering, commercial e-mail providers can identify communities of users more easily \cite{kleinberg/lawrence:2001} and focus marketing more efficiently. In general, communication by e-mail can be interfered with as well as utilized more extensively due to the non-trivial topological features of the e-mail network that we found here. Exploring the web of e-mails does not only extend our knowledge of social and communication networks but it also shows how vulnerable and exploitable these systems can be.

In conclusion, we have shown that an e-mail network, where nodes are given by e-mail addresses and links by exchanged messages, exhibits both small-world properties and scale-free behavior. The e-mail network is studied in terms of undirected, as well as directed networks. Spreading of e-mail viruses is considered, based on the appropriate viewpoint of an undirected graph. The scale-free nature of the e-mail network strongly eases persistence and propagation of e-mail viruses but also points to effective countermeasures.

\begin{acknowledgements}
We thank  A.-L.\ Barab{\'{a}}si, J.\ Davidsen, S.\ N.\ Dorogovtsev, and A.\ Vespignani for useful discussions and comments. H.~E.\ acknowledges support by the Studienstiftung des deutschen Volkes (German National Merit Foundation). 
\end{acknowledgements}
\vspace{2cm}
\setcounter{section}{0}

\end{document}